\documentclass[twoside]{article}
\usepackage{fleqn,espcrc2}
\usepackage{graphicx}
\usepackage[figuresright]{rotating}

\newcommand{\as}  {\mbox{$\alpha_s$}}
\newcommand{\asmz}  {\mbox{$\alpha_s(M_{\mathrm Z})$}}

\newcommand{\be}   {\begin{equation}}
\newcommand{\ee}   {\end  {equation}}
\newcommand{\bea}  {\begin{eqnarray}}
\newcommand{\eea}  {\end  {eqnarray}}
\newcommand{\mbmz} {{\ifmmode m_b(M_{\mathrm Z})\else $m_b(M_{\mathrm Z})$\fi}}
\newcommand{\mbmu} {{\ifmmode m_b(\mu)\else $m_b(\mu)$\fi}}
\newcommand{\gevcc}{{\ifmmode \mathrm{GeV}/c^2\else GeV/$c^2$\fi}}
\newcommand{\mevcc}{{\ifmmode \mathrm{MeV}/c^2\else MeV/$c^2$\fi}}

\newcommand{\AmS}{{\protect\the\textfont2
  A\kern-.1667em\lower.5ex\hbox{M}\kern-.125emS}}

\title{$m_b$ from ALEPH and comparison with other LEP experiments.}

\author{Fabrizio Palla\address[CERN]{CERN, CH-1211 Geneva 23, Switzerland}
        \thanks{On leave from Istituto Nazionale di Fisica Nucleare,
        Sezione di Pisa}}
       
\begin{document}

\begin{abstract}
I will review the measurements of the $b$-quark mass performed by ALEPH and DELPHI. A large set of observables has been used together with detailed studies on jet algorithms. Very clear effects due to the $b$-quark mass running are observed. Comparing with the determinations at the $\Upsilon(4S)$ threshold, the measurements of the running $b$-quark mass at the $Z$ pole are consistent with the predicted evolution from QCD.
\vspace{1pc}
\end{abstract}

\maketitle

\section{Introduction}

The $b$-quark mass is one of the fundamental parameters of the QCD Lagrangian.
However, due to confinement, quarks do not appear as free particles and therefore the definition of their mass is ambiguous. In fact, quark masses can  either be defined as for free particles (such as the leptons) as the position of the pole of the propagator, or they can be interpreted as effective coupling constants in the Lagrangian. In the former definition the mass is called ``pole mass'' and does not run with energy; in the latter the mass is called ``running mass''; as such it is a scale and renormalization scheme dependent quantity. 

The $b$-quark mass is important for phenomenology since it enters many theoretical predictions of physical quantities such as CKM matrix elements, $b$-hadron inclusive semileptonic decays, total widths.

Most of the $b$-quark mass  determinations have been extracted at the $\Upsilon(4S)$ threshold, from the $\Upsilon$ bound states with QCD sum rules \cite{sum_rules} and QCD lattice calculations\cite{lattice}. It is therefore interesting to measure this parameter at higher scales such as the one offered by LEP.

At LEP, the possibility to test the running of the $b$-quark mass within the framework of the perturbative QCD had not been  considered until very recently. The reason is that the effects of the mass rapidly become  very small with increasing energy for many observables, since they are proportional to $m_b^2/M_Z^2$ ($\cal O$ (0.1\%)), for instance in the total cross section. For other  quantities, such as jet rates, the effects are of the order of few percent due to
the suppression of the radiation off $b$ quarks.
 The advent of theoretical calculations at next-to-leading order (NLO) in QCD perturbation theory allowed for the determination of the $b$-quark mass~\cite{Rodrigo}.

In this article I will review the measurements done by ALEPH\cite{ALEPH_paper} using a large set of event shape observables and by DELPHI\cite{DELPHI_paper1,DELPHI_paper2} using the three-jet rate. The different  sensitivity to the hadronization corrections and next-to-leading order corrections can be used to investigate the interpretation of the results.

\section{Analysis Method}
        \label{method}
The method to extract the $b$-quark mass is based on measuring the ratio of any infrared safe observable $O$ computed for  $b$ and $uds$ (referred as $q$) induced events and assuming $\alpha_s$ universality,
\be
 R_{bq}^{pert} = O_b/O_q.
\ee

The following variables have been studied by ALEPH:
\begin{itemize}
\item The rate of  three-jet events, where jets are reconstructed using the {\tt DURHAM} \cite{Durham} clustering algorithm with $y_{cut}=0.02$.
\item The first and the second moments of the event shape variables Thrust T, C-parameter C, $y_3$, Total and Wide Jet Broadening, $B_T$ and $B_W$. The 
definitions of the observables can be found, e.g., in 
\cite{QCDmega,CataniBroad} and references therein.
\end{itemize}

DELPHI has used the three and the four jet event rates. 

The measured ratio $R^{\mathrm{meas}}_{bq}$ of the observables in $b$-tagged
events and in $uds$-tagged events 
can be related to the
quantities at the parton level
 via the following formula
\begin{eqnarray}
 \label{correction}
  R^{\mathrm{meas}}_{bq} =
\left(O_bH_bD_bT_{bb}{\cal P}_{bb}\right. +&
      O_cH_cD_cT_{bc}{\cal P}_{bc}+\nonumber\\  \left. 
  O_qH_qD_qT_{bq}{\cal P}_{bq}\right)/ &
 \left(O_bH_bD_bT_{qb}{\cal P}_{qb}+\right.\nonumber\\
  O_cH_cD_cT_{qc}{\cal P}_{qc}+& 
\left. O_qH_qD_lT_{qq}{\cal P}_{qq}\right). 
\end{eqnarray}
Here $H_x, D_x, T_{yx}$ are the corrections due to hadronization,
detector effects and tagging, respectively, and 
${\cal P}_{yx}$ are the purities of the tagged sample, where
$x$ is the true flavour and $y$ is the tagged one.
 The tagging corrections take into
account biases introduced by the flavour tag.
In case of ALEPH the ratio is computed with respect to an unbiased sample of all flavours, which
means that ${\cal P}_{qx}$ is equal to the ratio of the partial width of the  
$Z$ to $x$ quarks and the total hadronic width, and $T_{qx}=1$.

All the correction factors and purities are obtained from
Monte Carlo (MC) simulations. 
Because mainly ratios of corrections are involved,
some systematic uncertainties cancel.

$R^{\mathrm{pert}}_{bq}$ is extracted from the relationship (\ref{correction})
and finally corrected for the contribution of anomalous triangle
diagrams \cite{triangle}, in order to relate $R^{\mathrm{pert}}_{bq}$ to
$R^{\mathrm{pert}}_{bd}$ for which the perturbative calculations have been
performed. 

The general form of the NLO prediction for $R_{bd}^{pert}$ as a function of
the runninng $b$-quark mass $m_b$ is of the form
\begin{equation}
\label{r_bmass}
R_{bd}^{pert} = 1+\frac{m_b^2}{M_Z^2}\left[ b_0(m_b) + \frac{\alpha_s}{2\pi}b_1(m_b)\right].
\end{equation}
The ratio can also be expressed in terms of the pole mass $M_b$
\begin{equation}
\label{r_bmasspole}
R_{bd}^{pert} = 1+\frac{M_b^2}{M_Z^2}\left[ b^P_0(M_b) + \frac{\alpha_s}{2\pi}b^P_1(M_b)\right].
\end{equation}

The two predictions are equivalent at this order. The
coefficient functions $b_{0,1}$ for the two schemes can be related to each other
by expressing the pole mass in terms of the running mass
\be
 \label{poletorun}
    M_b = m_b(\mu) \left[ 1 + \frac{\as(\mu)}{2\pi}
            \left( \frac{4}{3} - 2 \ln\frac{m_b(\mu)^2}{\mu^2} \right)
                     \right] .
\ee

The coefficient functions $b_0$ and $b_1$ have been computed for the ratio 
of three
and four jet rates  \cite{Rodrigo,zbb4}. The three jet
rate ratio allows a NLO prediction, while for the four jet rate only a LO
prediction is available. For the other variables used by ALEPH the
predictions have been obtained using the MC generators ZBB4\cite{zbb4}
and EVENT\cite{event} which are correct to NLO.
The hadronization corrections have been evaluated for all variables by
computing the relevant observables at parton and at hadron level.

\section{ALEPH analysis}

In the ALEPH analysis data taken at the peak of the Z resonance
from 1991 to 1995 are used.
A standard hadronic event selection \cite{QCDmega} is applied,
which is based on charged particles. A
cut $|{ \cos\theta_{T} }| < 0.7$ is imposed, where
$\theta_{T}$ is the polar angle of the thrust axis, computed
from all charged and neutral particles as obtained from the 
energy-flow algorithm \cite{EFLOW}. This requirement ensures that
the events are well contained within the vertex detector (VDET) acceptance.
 After the selection, a sample of 2.3 million
hadronic events remains for further analysis, with about 0.3\% of non-hadronic
background, mainly coming from  $\tau^+\tau^-$ events.

An important ingredient of the analysis is the choice of 
the Monte Carlo simulation, which is based on on 
 {\tt JETSET 7.4} parton shower
model plus string fragmentation \cite{Jetset}. 
The production rates, decay modes and lifetimes
of heavy hadrons are adjusted to agree with recent measurements,
while heavy quarks are fragmented using the Peterson et al.\ model 
\cite{Peterson}.
\begin{table*}[htb]
  \caption{      
Leading (LO) and next-to-leading (NLO) order contributions to $1 - R^{\mathrm{pert}}_{bd}$, for the running mass (run) and the pole mass (pol) schemes. The contributions are evaluated for a running (pole) mass of 3 (5) \gevcc. The strong coupling \asmz\ is set to 0.119. The values are given for the three-jet rate ({\tt DURHAM} scheme) and for the event-shape variables thrust $T$, $C$ parameter, the transition value $y_3$ for three to two jets ({\tt DURHAM} scheme), and the total and wide jet broadenings ($B_T$ and $B_W$). The indices indicate the first or second moment of the event shape variable.}
\label{lo-nlo:tab}
\newcommand{\cc}[1]{\multicolumn{2}{c}{#1}}
\renewcommand{\tabcolsep}{2pc} 
\renewcommand{\arraystretch}{1.2} 
\begin{center}
\begin{tabular}{@{}crrrr}
      \hline
          & \cc{run} & \cc{pol} \\
      $O$ & LO            & NLO           & 
            LO            & NLO              \\
      \hline
      \rule{0pt}{5.5mm}%
 $R_3      $&  $0.020$  &   $0.010$  &  $0.056$  &  $-0.008$   \\
      $T_1$ &  $0.036$  &   $0.019$  &  $0.076$  &  $-0.007$   \\
      $T_2$ &  $0.017$  &   $0.032$  &  $0.043$  &  $ 0.036$   \\
      $C_1$ &  $0.044$  &   $0.022$  &  $0.091$  &  $-0.011$   \\
      $C_2$ &  $0.021$  &   $0.039$  &  $0.052$  &  $ 0.043$   \\
  $y_{3_1}$ &  $0.032$  &   $0.007$  &  $0.071$  &  $-0.021$   \\
  $y_{3_2}$ &  $0.015$  &   $0.003$  &  $0.032$  &  $-0.007$   \\
  $B_{T_1}$ &  $0.117$  &   $0.006$  &  $0.188$  &  $-0.074$   \\
  $B_{T_2}$ &  $0.036$  &   $0.112$  &  $0.080$  &  $ 0.123$   \\
  $B_{W_1}$ &  $0.117$  &  $-0.085$  &  $0.188$  &  $-0.183$   \\
  $B_{W_2}$ &  $0.036$  &   $0.016$  &  $0.080$  &  $-0.013$   \\
    \hline
    \end{tabular}
\end{center}
\end{table*}

The observables 
described in Section \ref{method} are computed using only
charged tracks. 
    
In the analysis only $b$ events are tagged. The selection is based on
the long $b$ hadron lifetime thanks to the great precision of the VDET,
 and follows closely the approach described in detail in
Ref.~\cite{Rb}.
The actual  selection of $b$ events is obtained from a cut on the
distribution of the confidence level $P_{uds}$ 
that all tracks of the event come from the main vertex. 
Very good agreement between data and MC is observed. 
On  $b$ events the selection has an efficiency 
of $80.5\%$ and a purity of ${\cal P}_{bb} = 83.1\%$. The 
backgrounds amount to ${\cal P}_{bc} = 13.5\%$ and ${\cal P}_{bq} = 3.4\%$.

        \subsection{Choice of variables}

A list of the leading (LO) and 
next-to-leading order (NLO) contributions to $1 - R^{\mathrm{pert}}_{bd}$ is
given in Table \ref{lo-nlo:tab} for all variables, 
both for the running and the pole mass schemes. 
They have been evaluated for a $b$-quark mass of 3 \gevcc\ in the former and 5 \gevcc\
in the latter case. 
It is found that for some observables such as the jet
broadening variables the mass effect is rather large. However,
the NLO corrections can also be  sizeable, as in the case of the second moment
of thrust and the first moment of the wide jet broadening. For such 
variables it would definitely be necessary to also compute 
the NNLO contributions
in order to obtain a reliable perturbative prediction. 
Because of these observations only those 
variables are considered for which in both schemes the NLO contribution
is clearly smaller than the LO term.
\subsection{Hadronization corrections}
 \label{hadcorr}

The perturbative predictions are corrected for hadronization effects
by computing the relevant observables at parton and at hadron level,
including final state photon radiation off quarks.
Several Monte Carlo models based on the parton shower approach
plus subsequent string or cluster fragmentation are employed for
this purpose. For the nominal analysis, the same generator as for the
full simulation is used.

The ratios of hadronization corrections, $H_{b/q}$,
are listed in Table \ref{statres:tab}. The 
corrections are rather sizeable for almost all the observables; in most
cases they are of the same size as or larger than the expected mass effect.
Only the three-jet rate and the first two moments of the $y_3$
distribution have corrections at the 2\% level or below. 
For all the other event-shape variables the corrections are of 
the order of 10\% or even larger. 
It has been found that the deviation from unity
is almost entirely due to $b$ hadron decays, which change the distributions
mainly in the two-jet region. As can be observed from Table \ref{statres:tab},
the same corrections, computed taking only hadrons which stem directly
from the string before any decay, come close to unity within one or
two percent. This is in agreement with expectations from recent 
calculations of nonperturbative power-law corrections to moments
of event-shape distributions for massive quarks \cite{Trocsanyi:2000ta}.

By requiring that the hadronization corrections do not exceed 2\% two
variables remain: the first moment of $y_3$ and the three jet ratio.

\begin{figure}[htb]
  \begin{center}
    \includegraphics[width=\linewidth]{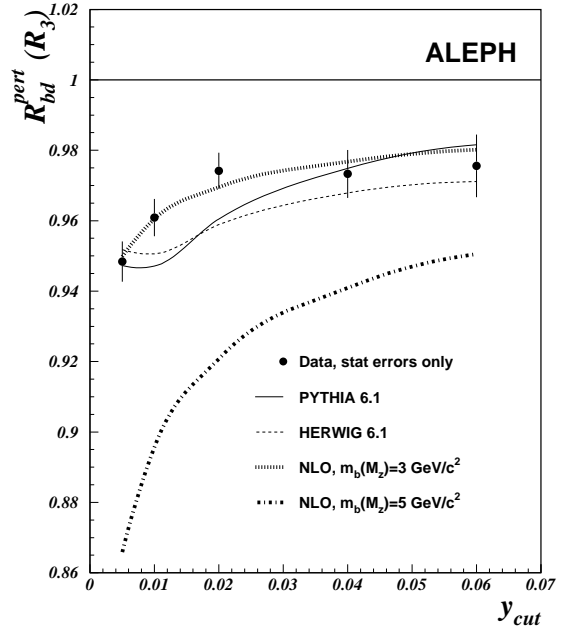}
      \caption{Comparison of the $y_{\mathrm{cut}}$ dependence of the measured
    ratio $R^{\mathrm{pert}}_{bd}$ for the three-jet rate 
    to the predictions of parton shower models
    as well as next-to-leading order (NLO) perturbative QCD for two
    different values of the 
    $b$-quark mass in the $\overline{\mathrm{MS}}$ scheme.
     The errors are statistical only. \label{fig:r3bd_ycut}}
  \end{center}
\end{figure}
\begin{figure}[htb]
  \begin{center}
      \includegraphics[width=\linewidth]{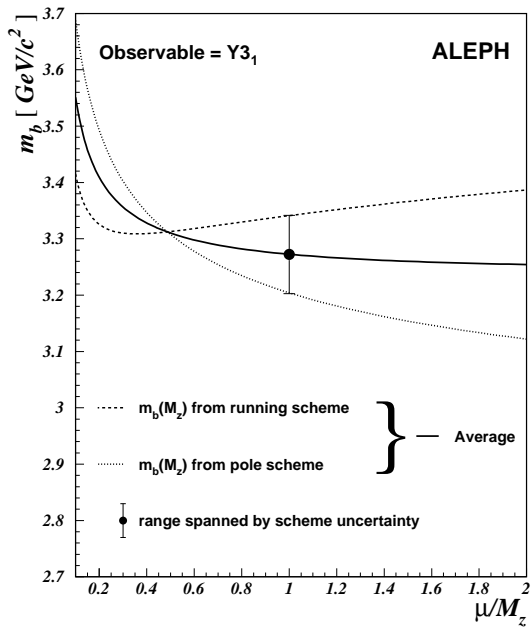} 
    \caption{    Renormalization scale dependence for the extracted $b$-quark
    mass in the running and pole mass scheme, for the observable $y_{3_1}$. 
    The error bar spans the range 
    of the scheme uncertainty. The full line indicates the renormalization
    scale dependence of the average computed from the 
    running and pole mass schemes.
    \label{mudep:fig}}
  \end{center}
\end{figure}

\begin{table*}[htb]
    \caption{
             ALEPH results for $R^{\mathrm{pert}}_{bd}$ with statistical
             errors, including the uncertainty from the MC statistics.
             The results are obtained with the hadronization 
             corrections $H_{b/q}$ as predicted by the MC. Also given are
             the corrections $H^{nod}_{b/q}$, computed from hadrons
             directly originating from the string, before any decays
             (\textit{nod}=no decays).
            }
\renewcommand{\tabcolsep}{2pc} 
\renewcommand{\arraystretch}{1.2} 
  \begin{center}
    \begin{tabular}{@{}cccc}
      \hline
      $O$ & $R^{\mathrm{pert}}_{bd}$ & $\;H_{b/q}\;$ & $\;H^{nod}_{b/q}\;$ \\
      \hline
      \rule{0pt}{5.5mm}%
             $R_3$ &  $0.974\pm0.005$& 0.980  &0.987  \\
             $T_1$ &  $0.910\pm0.002$& 1.134  &1.003  \\
             $C_1$ &  $0.894\pm0.002$& 1.169  &1.009  \\
         $y_{3_1}$ &  $0.955\pm0.005$& 1.023  &0.987  \\
         $y_{3_2}$ &  $0.979\pm0.010$& 0.984  &0.984  \\
         $B_{T_1}$ &  $0.831\pm0.001$& 1.306  &1.025  \\
         $B_{W_2}$ &  $0.929\pm0.003$& 1.088  &0.986  \\
    \hline
    \end{tabular}     
  \end{center}
\label{statres:tab}
\end{table*}

\subsection{Measurement of \boldmath$R^{\mathrm{pert}}_{bd}$\unboldmath}
 \label{rmeas}

  The results for the perturbative ratio
 $R^{\mathrm{pert}}_{bd}$ are listed in Table \ref{statres:tab}, after having applied the
correction procedure defined in equation~\ref{correction} and correcting for anomalous triangle
diagrams.

The dependence on the resolution parameter $y_{\mathrm{cut}}$
of the perturbative ratio $R^{\mathrm{pert}}_{bd}$ for the three-jet rate
is indicated in Fig.~\ref{fig:r3bd_ycut}.

The data are compared to the predictions of the parton shower
models of {\tt PYTHIA 6.1}, which is based on {\tt JETSET},
and {\tt HERWIG 6.1} \cite{Herwig}. The MC models
are in reasonable agreement with
the measurement at large $y_{\mathrm{cut}}$ values, where the statistical
uncertainty is large, however. At intermediate resolution
parameters these models predict a lower ratio than observed.

In addition, in Fig.~\ref{fig:r3bd_ycut} the next-to-leading order perturbative
QCD predictions for two different values of the running $b$-quark mass
in the $\overline{\mathrm{MS}}$ scheme are given.
The data clearly favour a mass close to 3 \gevcc.

        \subsection{Systematic errors}
\label{systematics}
Systematic uncertainties on the mass extraction can be divided into three categories: the ones coming
from uncertainties on the tagging biases, from the hadronization and from
the missing higher order corrections.

The first ones have been evaluated by varying  the gluon splitting rate which
influences the purity of the $b$ tag and by studying the variation of
the results due to the inadequacy of the simulation of the tracking detectors.

The uncertainties from the modelling of the hadronization are typically
evaluated by computing the hadronization corrections with different MC 
generators.
As has been shown in Section \ref{hadcorr}, the $b$ hadron decays have
a large impact on the size of the hadronization corrections.
Because in the tuning of the MC hadronic final states are analyzed
after all decays, differences in the description of hadron decays can
lead to differences in the tuned fragmentation parameters.
In order to assess an uncertainty related to the string fragmentation 
parameters and subsequent decays,
the variation in the final result when using two different $b$ hadron 
decay tables
for the
hadronization corrections is taken as a systematic uncertainty.

Another quantity relevant for this analysis 
is the $b$ fragmentation function, which describes the
energy fraction transferred to the $b$ hadrons during the fragmentation
process. Three different sets of different fragmentation functions were
used to assess the systematic uncertainty due to this source.
Finally, in order to study
purely the difference between string and cluster fragmentation, 
the variation in the hadronization corrections for light quarks ($uds$)
only has been propagated into an uncertainty  on the final result.

\begin{figure}[htb]
  \begin{center}
      \includegraphics[width=\linewidth]{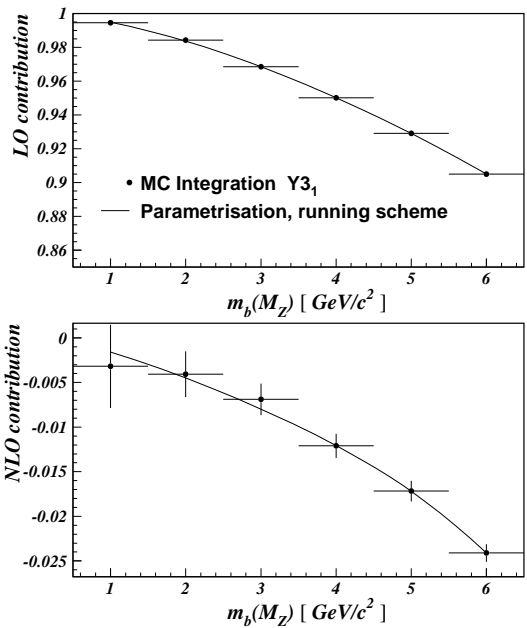}
    \caption{
    Parametrisations of the $b$-quark mass dependence for the 
    LO (top) and NLO (bottom) contributions to $R^{\mathrm{pert}}_{bd}$
    in the running mass scheme for
    the first moment of the $y_3$ distribution. The points indicate the result
    of the MC integration, and the full line the parametrisations. 
    The NLO contributions are evaluated using 
    $\alpha_s(M_{\mathrm Z})=0.119$.
    \label{param_thr:fig}}
  \end{center}
\end{figure}

Missing higher order corrections are estimated by extracting firstly the
pole mass from the perturbative expression in the pole mass scheme,
then translating that result into a running quark mass $m_b(m_b)$ at the
$b$-mass scale, and finally running this mass up to the $M_{\mathrm Z}$ scale
using the renormalization group equation.
In addition, the effects of uncalculated higher order terms can be estimated by
a change in the renormalization scale. For the central value
$\mu = M_{\mathrm Z}$ is employed, and the systematic error is taken to be half of the range
of mass values (average of running and pole mass schemes)
found when varying $\mu$ from 
$0.1 M_{\mathrm Z}$ to $2 M_{\mathrm Z}$. 
The behaviour of the extracted mass in the two schemes as a function of the renormalization scale $\mu$ is illustrated in Fig.~\ref{mudep:fig} for the first moment of the $y_3$ variable.

\subsection{Results for \boldmath$\mbmz$\unboldmath}
 \label{resultbmass}

Based on the predictions obtained above, the running $b$-quark mass is
determined from the measured ratio $R^{\mathrm{pert}}_{bd}$ for two
observables, the first moment of the $y_3$ distribution and the
three-jet rate.

As stated in Section \ref{method}, the functional dependence of the coefficients $b_{0,1}$ for the 
moments of the event shape variables are computed by the ratio of the cross sections
for $b$ and $d$ quarks using the  MC generators 
ZBB4 \cite{zbb4} and EVENT \cite{event}. ZBB4
allows for the integration of
the fully differential NLO matrix elements including
mass effects, whereas EVENT contains the massless expressions \cite{ERT}.

In Fig.~\ref{param_thr:fig} an example of these fits is shown for
the first moment of the $y_3$ distribution. 
With these parametrisations it is possible to estimate the
actual quark mass effects in leading order  and its next-to-leading order
corrections. In Table \ref{lo-nlo:tab} 
a list of these LO and NLO contributions is
given for all variables for both the running and the pole mass schemes.

The results are listed in Table \ref{fullmassres:tab} together with
the statistical and systematic uncertainties, which have been propagated
from the corresponding uncertainties on $R^{\mathrm{pert}}_{bd}$. 
The result from the first moment of the $y_3$ observable 
is quoted as the final $b$-quark mass value, due to its smaller hadronization and 
systematics errors.
The theoretical uncertainty is estimated as described in 
Section \ref{systematics} by evaluating the impact of the 
uncertainty on the strong coupling constant and the renormalization scale 
variation.
Because the pole and the running mass scheme 
are equivalent at NLO, the average of the values found for the two 
schemes is quoted as the final result, 
and half of the difference is taken as an additional theoretical systematic uncertainty due 
to the scheme ambiguity.
This leads to a measurement of the $b$-quark mass from ALEPH of
\begin{eqnarray}
  m_b(M_{\mathrm Z})  =  \left[ 3.27 \,\pm\, 0.22 (\mathrm{stat}) \,\pm\, 0.22 (\mathrm{exp})\right. 
\nonumber\\
\left. \,\pm\, 0.38 (\mathrm{had})  \,\pm\, 0.16 (\mathrm{theo})\right] \;\gevcc.  
\end{eqnarray}

\begin{table*}[htb]
    \caption{
             ALEPH measured $b$-quark mass $m_b(M_{\mathrm Z})$ in the 
             $\overline{\mathrm{MS}}$ scheme with statistical
             (stat), experimental (exp) and hadronization (had)
             uncertainties. 
             In the last column the measured
             pole mass $M_b$ is listed.
            }
  \vspace{0.3cm}
  \begin{center}
    \begin{tabular}{@{}cccccc}
      \hline
      $O$ & $m_b(M_{\mathrm Z})\, [\gevcc]$ & 
          $\pm$(stat)& 
          $\pm$(exp) & 
          $\pm$(had) & 
          $M_b\, [\gevcc]\;$ \\
      \hline
      \rule{0pt}{5.5mm}%
        $R_3$ & 2.76 & 0.28 & 0.28 & 0.62 & 3.65 \\
    $y_{3_1}$ & 3.34 & 0.22 & 0.22 & 0.38 & 4.73 \\
    \hline
    \end{tabular}     
  \end{center}
\label{fullmassres:tab}
\end{table*}
\section{DELPHI analysis}

The first measurement of the running $b$-quark mass at scales of the $Z$ mass
has been performed by DELPHI, using 1992 to 1994 data and the three jet rate ratio
using the Durham algorithm for jet clustering \cite{DELPHI_paper1}.
Effects of the running of the $b$-quark mass have also been measured in the four jet rate ratio as 
reported in \cite{Garcia}.

A new analysis has been presented at this conference which is based on the 1994 and 1995 data sample, which amounts to about 1.3 Million hadronic $Z$ decays.
A better $b$ tagging algorithm was developed, which is described in \cite{Delphi_btag}. The efficiency to tag $b$ events is about 53\% with a 85\% purity.

The Monte Carlo model used is based on {\tt PYTHIA}, with the DELPHI tuning.

The study was performed using the {\tt DURHAM} and {\tt CAMBRIDGE}\cite{Cambridge} jet algorithms. The latter  is found to  minimize the NLO corrections to the three jet rate for the running $b$-quark mass~\cite{Cambridge}. 

        \subsection{Measurement of \boldmath$R^{\mathrm{pert}}_{bd}$\unboldmath}

\begin{figure*}[htb]
  \begin{center}
        \begin{tabular}{@{}cc}
      \includegraphics[width=0.5\linewidth]{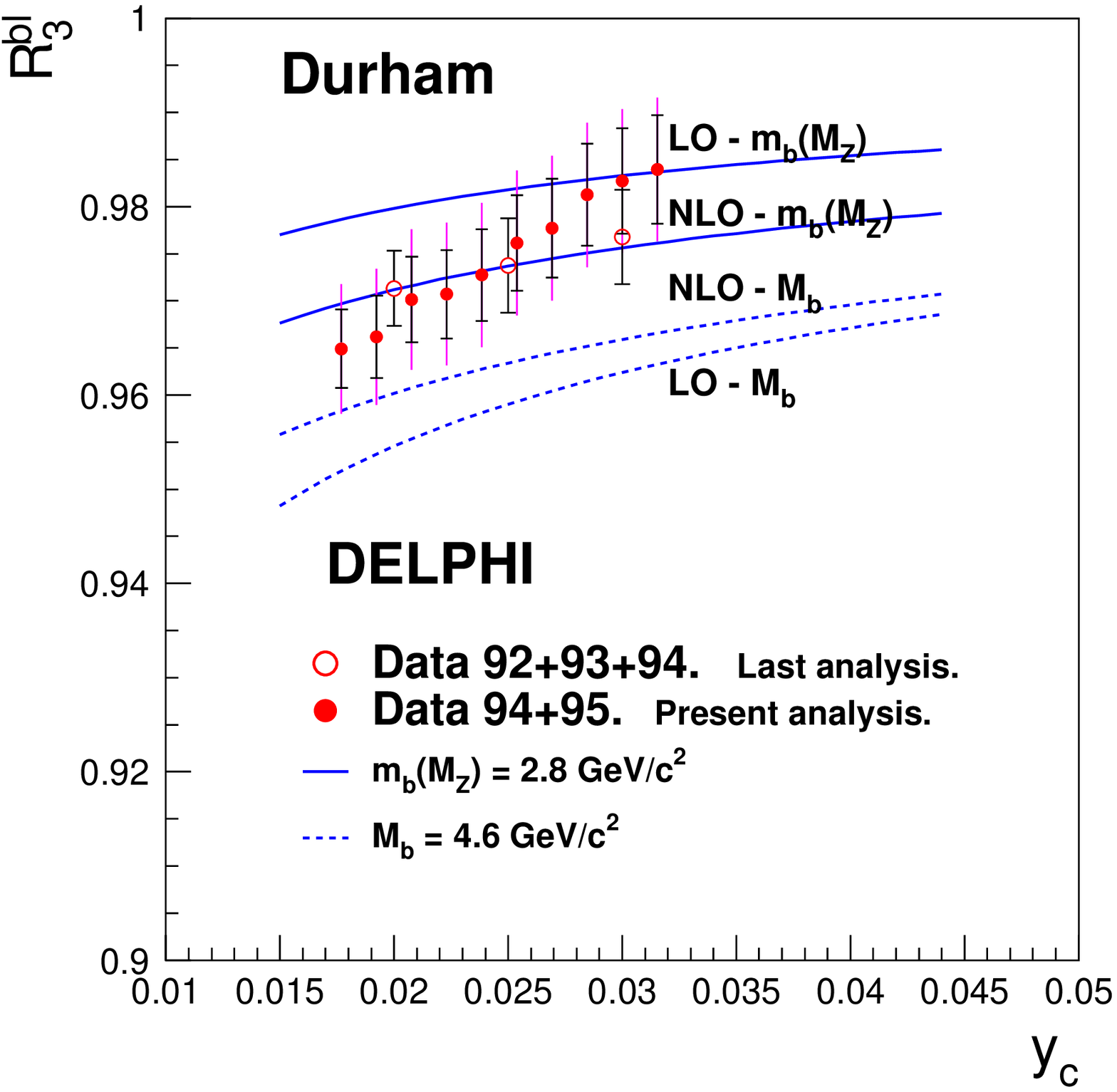} &
      \includegraphics[width=0.5\linewidth]{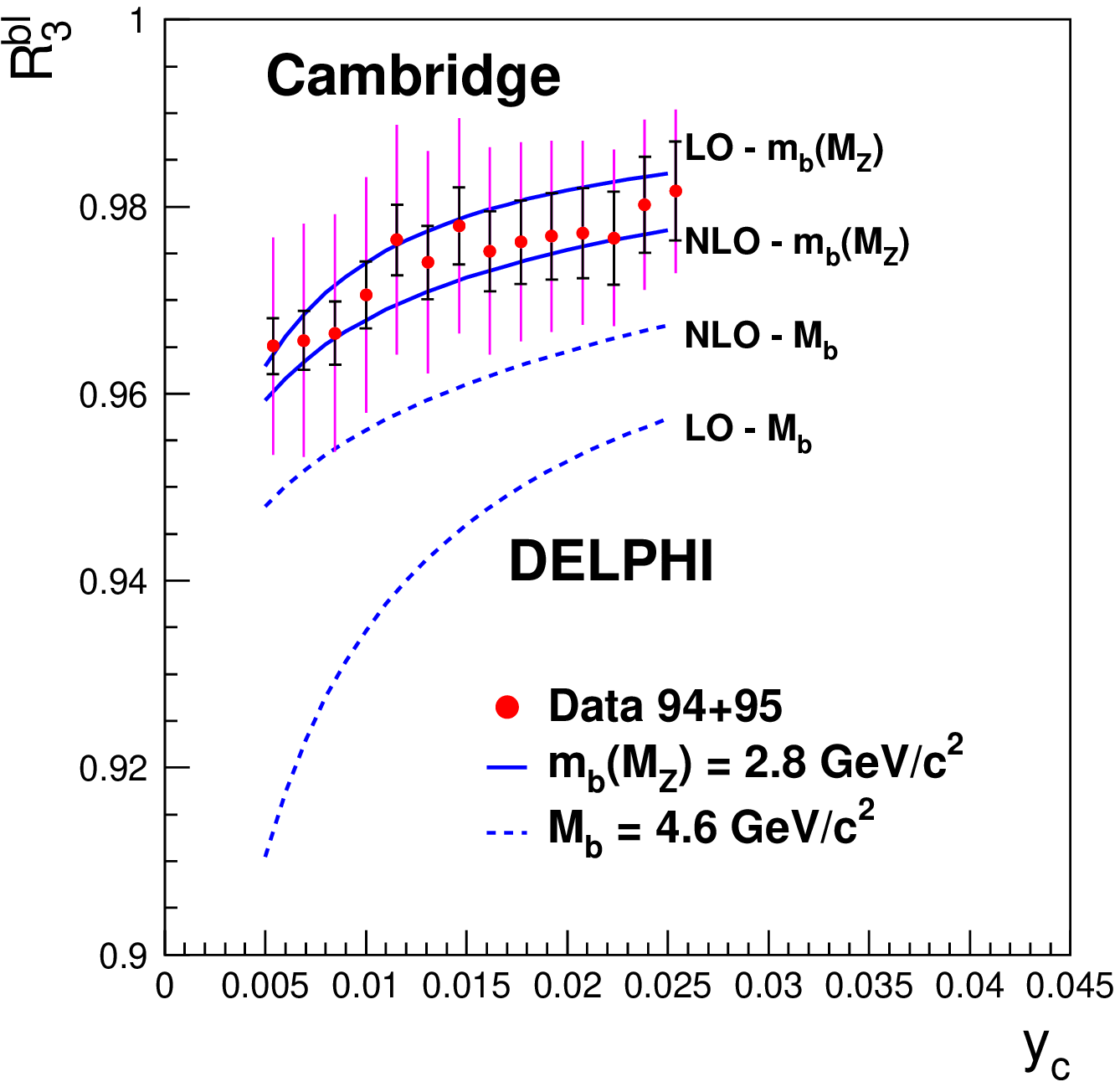} 
        \end{tabular}
    \caption{Corrected data values of the perturbative three jet ratio using {\tt DURHAM} (left) and {\tt CAMBRIDGE} (right) jet algorithms for DELPHI, compared with the theoretical predictions from \cite{Rodrigo} at LO and NLO in terms of the pole mass $M_b=4.6$~\gevcc (dashed lines) and in terms of the running mass \mbmz$=2.8$~\gevcc (solid lines).
    \label{Delphi_r3}}
  \end{center}
\end{figure*}

The three jet ratio  is shown in Figure \ref{Delphi_r3} for both jet 
algorithms. As can be seen the LO calculation did not provide good description of the data and NLO corrections are indeed needed. In the case of  {\tt DURHAM} the data are closer to the running $b$-quark mass calculations, despite the NLO corrections are larger with respect to the 
pole mass. For the {\tt CAMBRIDGE} algorithm data still prefer the running $b$-quark mass calculations, but here the NLO corrections are in fact smaller for this scheme.

        \subsection{Systematic errors}

\begin{figure}[htb]
  \begin{center}
      \includegraphics[width=\linewidth]{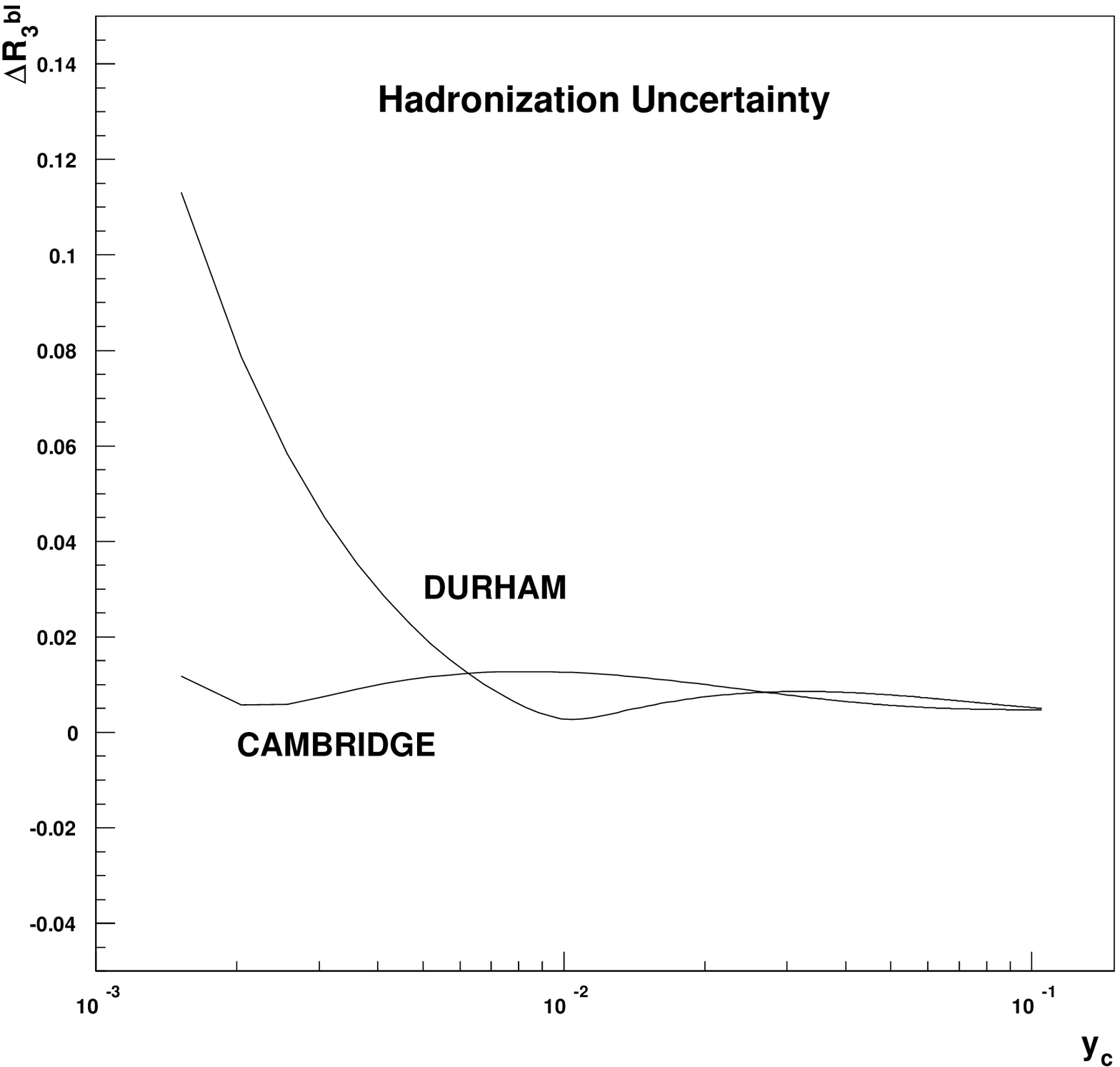} 
    \caption{Evolution of the total hadronization error on the three jet ratio for the DELPHI analysis with the resolution parameter $y_c$. The error corresponds to two times the sum in quadrature of $\sigma_{tun}$ and  $\sigma_{mod}$, as described in the text.}
    \label{had_err}
  \end{center}
\end{figure}

Hadronization errors were derived from comparisons with different Monte Carlo models tuned to describe the data. Two different sources of errors were considered: $\sigma_{mod}$ which describes the difference of the hadronization corrections between {\tt PYTHIA} and {\tt HERWIG}, and $\sigma_{tun}$ which accounts for the possible variation of the main fragmentation parameters in {\tt PYTHIA}.
The first uncertainty takes into account both the different fragmentation scheme (string or cluster) and the different $b$ hadron decay tables.
In the case of  $\sigma_{tun}$ the error was evaluated by changing the most relevant fragmentation parameters of the string model in {\tt PYTHIA} by $\pm 2\sigma$ from their central tuned values and assuming they are uncorrelated.
The two above errors were combined in quadrature to estimate the hadronization uncertainty. 
The interval of validity for the jet resolution parameter $y_{cut}$ to perform the measurement was chosen to have small hadronization error (below 5\%) and smooth $y_{cut}$ dependence. Figure \ref{had_err} shows the total hadronization 
correction uncertainty for the three jet rate for the two jet algorithms. 

The uncertainty coming from the tagging is taken into account by varying the purities by 1\%.

The errors coming from missing higher orders are estimated by varying the renormalization scale $\mu = M_{\mathrm Z}$ from $0.5$ to $2$. The ambiguity due to the definition of the three jet rate in terms of the running or the pole $b$-quark mass has been evaluated  as explained in Section~\ref{systematics}. 

        \subsection{Extraction of the $b$-quark mass}

The values for the running $b$-quark mass obtained with the {\tt DURHAM} and
 {\tt CAMBRIDGE} algorithms are shown in Table~\ref{Delphi:tab}, for $y_{cut}=0.02$ and $0.005$, respectively.

The three jet rate measured with the {\tt CAMBRIDGE} algorithm is in agreement with both LO and NLO calculations in terms of the running  $b$-quark mass. For this reason the mass is extracted from this algorithm using the NLO prediction. The value of the running $b$-quark mass from DELPHI is therefore
\begin{eqnarray}
        m_b(M_{\mathrm Z}) = \left[ 2.61 \,\pm\, 0.18  (\mathrm{stat}) \,
^{+0.45}_{-0.49} (\mathrm{frag.}) \right.\nonumber\\
\left. \,\pm\, 0.18 (\mathrm{tag.}) \,\pm\, 0.07  (\mathrm{theo.}) \right] \;\gevcc.
\end{eqnarray}
\begin{table*}[htb]
    \caption{
             DELPHI measured $b$-quark mass $m_b(M_{\mathrm Z})$ in the 
             $\overline{\mathrm{MS}}$ scheme with statistical
             (stat), hadronization (frag.), $b$ tagging (b-tag)
             and theoretical (theo)
             uncertainties. The first line contains the results
             of the previous published analysis.
            }
  \vspace{0.3cm}
  \begin{center}
    \begin{tabular}{@{}clcccccc}
      \hline
      $y_{cut}$ & Algorithm &$m_b(M_{\mathrm Z})\, [\gevcc]$ & 
          $\pm$(stat)& 
          $\pm$(frag) & 
          $\pm$(b-tag) & 
          $\pm$(theo)  & Data\\
      \hline
        0.02 & {\tt DURHAM} &$2.67$ & $0.25$ &$0.34$ & -    & $0.27$ &\cite{DELPHI_paper1}\\
        0.02 & {\tt DURHAM} &$2.81$ & $0.25$ &$0.34$ & $0.20$ & $0.27$ &\cite{DELPHI_paper2}\\
        0.005 & {\tt CAMBRIDGE} &$2.61$ & $0.18$ &$^{+0.45}_{-0.49}$ & $0.18$ & $0.07$ &\cite{DELPHI_paper2}\\\hline
    \end{tabular}     
  \end{center}
\label{Delphi:tab}
\end{table*}
\begin{figure*}[htb]
  \begin{center}
    \includegraphics[width=0.5\linewidth]{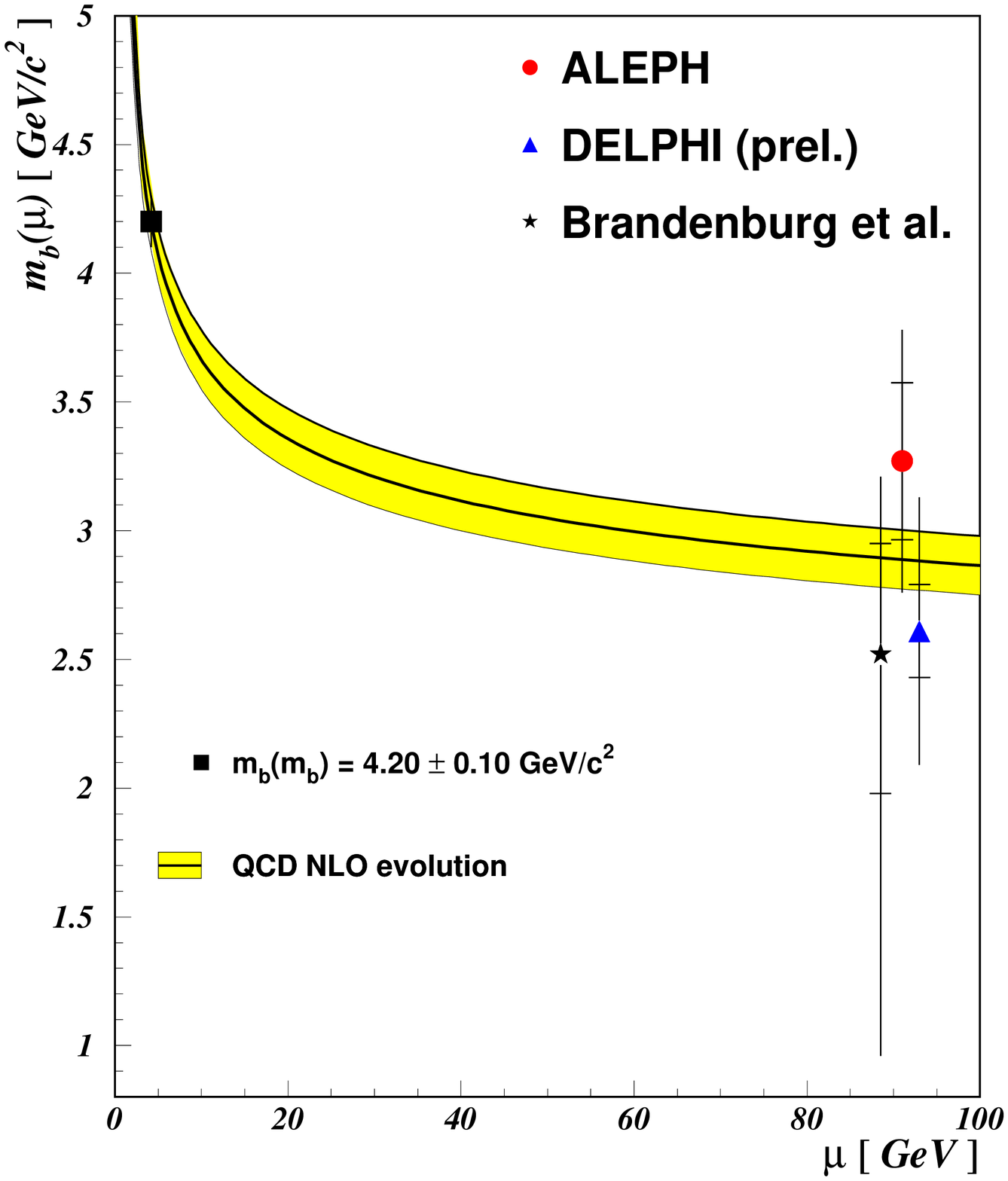}
    \caption{
     Comparison of the ALEPH and DELPHI result for $m_b(M_{\mathrm Z})$ 
     with the world average value
     of low-energy measurements for $m_b(m_b)$ \cite{Pich}, which is evolved up to
     the $M_{\mathrm Z}$ scale using a two-loop evolution equation with 
     $\alpha_s(M_{\mathrm Z})=0.119\pm0.003$. Also shown is the measurements
     by Brandenburg et al.\ \cite{SLACbm1}.
     The inner error bars indicate the quadratic sum of the statistical and
     experimental uncertainties. The three points at the Z pole are
     separated for clarity.
    \label{Bmass_run}}
  \end{center}
\end{figure*}

\section{Conclusions}

ALEPH has studied a large set of observables to extract the $b$-quark mass at the $Z$ peak. The first moment of $y_3$ has the smallest hadronization and systematic errors.  DELPHI used instead the three jet ratio using the {\tt CAMBRIDGE} jet algorithm which is found to  minimize the NLO corrections for the running $b$-quark mass. 

Clear effects due to the running of the $b$-quark mass are visible and confirm the expectation from QCD. The results from ALEPH, DELPHI and a measurement from Brandeburg {\it et al.} \cite{SLACbm1} are shown in Fig.~\ref{Bmass_run}, together with the average of the low energy measurements for $m_b(m_b)$ \cite{Pich}, which is evolved up to the $Z$ mass scale using a two-loop evolution equation with \as$=0.119\pm0.003$.

The main error on the measurement comes from the uncertainties affecting the  hadronization corrections. A big fraction of it is coming  from $b$ hadron decays as has been estimated in great detail by ALEPH. 

Despite of the fact that the $b$-quark mass measured from the three jet ratio from ALEPH and DELPHI shows
a good agreement.

 The mass extracted from the first moment of $y_3$ seems
to be larger and barely compatible with the jet ratio measurements. 

The spread in the central value and the actual size of the errors indicate that the limiting precision of the measurement of the $b$-quark mass at the $Z$ mass scale is of the order of $500$ \mevcc.

\section{Acknoledgements}

I would like to thank Prof. Narison for ruling  the Conference in a nicely informal atmosphere. I am also grateful J. Fuster, S. Mart\'{\i} i Garc\'{\i}a and G. Dissertori for useful discussions.

\end{document}